\documentclass[twocolumn,showpacs,preprintnumbers,amsmath,amssymb]{revtex4}\newcommand{\statevector}{\sva}

\usepackage{graphicx}
\usepackage{dcolumn}
\usepackage{bm}
\usepackage[mathscr]{eucal}
\usepackage[nooneline]{subfigure}
\newcommand{\etal}{\textit{et al}}
\newcommand{\qst}{quantum state transfer}

\newcommand{\microns}{$\mu$m}
\newcommand{\bra}[1]{\langle#1\vert}
\newcommand{\ket}[1]{\vert#1\rangle}

\newcommand{\braket}[1]{\langle #1 \rangle}

\newcommand{\eqn}[1]{Eq.~(\ref{#1})}
\newcommand{\eqns}[1]{Eqs.~(\ref{#1})}
\newcommand{\ignore}[1]{}
\newcommand{\ahat}[1]{\hat a_{#1}}
\newcommand{\commute}[2]{\left[#1,#2\right]}
\newcommand{\dr}{\Delta \rho}
\newcommand{\vd}{\mathcal{D}}
\DeclareMathOperator{\tr}{tr}
\newcommand{\fig}[1]{Fig.~\ref{#1}}

\newcommand{\sva}{
\begin{widetext}
	\begin{equation}
		\ket{\psi(t)}=c_g\ket{gg}\ket{00}+c_e\left\{\alpha_1(t)e^{-i
		\phi_1(t)}\ket{eg}\ket{00}+\alpha_2(t)e^{-i \phi_2(t)}\ket{ge}\ket{00}
		\vphantom{\frac{\beta_s(t)}{\sqrt{2}}}\right.
		+\left.\ket{gg}\left(\frac{\beta_s(t)}{\sqrt{2}}(\ket{01}+\ket{10})+
		\frac{\beta_a(t)}{\sqrt{2}}(\ket{01}-\ket{10})\right)\right\}.
	\end{equation}
\end{widetext}}

\begin{document}

\preprint{APS/123-QED}

\title{The Effect of Stochastic Noise on Quantum State Transfer}

\author{T.M. Stace}
\email{tms29@cam.ac.uk}
\author{C.H.W. Barnes}
\affiliation{
Cavendish Laboratory, University of Cambridge,  Madingley Road, Cambridge CB3 0HE, United Kingdom}

\date{\today}

\begin{abstract}
We consider the effect of classical stochastic noise on control laser pulses used in a scheme for transferring quantum information between atoms, or quantum dots, in separate optical cavities via an optical connection between cavities.  We develop a master equation for the dynamics of the system subject to stochastic errors in the laser pulses, and use this to evaluate the sensitivity of the transfer process to stochastic pulse shape errors for a number of different pulse shapes.  We show that under certain conditions, the sensitivity of the transfer to the noise depends on the pulse shape, and develop a method for determining a pulse shape that is minimally sensitive to specific errors.
\end{abstract}

\pacs{
03.65.Yz, 
03.67.Hk, 
42.50.Lc, 
02.60.Pn 
}

\keywords{quantum communication, classical noise, decoherence, optimisation}
\maketitle

\section{Introduction} \label{Intro}

The ability to transfer a quantum state from one system to another will be a useful tool in the future of quantum information, both for quantum communication and for building large quantum computers from smaller components. Various schemes have been proposed for transferring quantum states \cite{cir97,pel97,enk97a,enk97,enk99,vri00,dua01}, some of which are based on teleportation and 
entanglement purification.  Others are more direct, having a quantum channel (typically an optical fiber) connecting the two systems, and classical control pulses applied to each system ensure that emission from the first system and absorption by the second are perfect.

Over long distances, it appears that protocols based on teleportation and purification are necessarily more efficient than direct schemes \cite{dua01}, but over short distances (e.g. for interconnects between quantum components), the more direct method, first proposed by Cirac \etal. \cite{cir97}, may be practical, and is probably easier to implement since it does not require quantum measurement.  The scheme by Cirac \etal.\ \cite{cir97}   implements \qst\ from one  atom located inside an optical cavity to another atom in a separate but connected cavity, mediated by Raman transitions via an atomic state (outside the computational basis) as well as cavity and external photons, and is controlled by laser pulses incident on each atom.  The transfer is therefore subject to amplitude and phase errors in the control pulses.

The idea \cite{cir97} is also applicable to single-electron spin based systems in semiconductors, with additional uses such as spin
measurement \cite{ima99,ima00}, and as a controllable source of decoherence for
initial state preparation.  The intermediate state required for spin-flip Raman
transitions between single electron spin states is a state known as a
trion \cite{ima99,ima00}.  This state is a bound state of two electrons and a single
hole, and couples to the single electron spin states via $x$- and $y$-polarised light. 
Physically, this corresponds to the optical excitation of an exciton which may form a
bound state with the electron, very much akin to an H$^-$ complex.  Trion states have
been observed experimentally in GaAs \cite{shi95, shi01}, and are relatively long 
lived, and so the scheme has the potential to be realised in semiconductor systems.

In this paper we present an analysis of the effect of two possible types of
stochastic error in control pulses, and consider a method for reducing the
sensitivity of the pulse to such errors.  Our results show that it is
possible to design pulse shapes that reduce the sensitivity of the transfer process to
certain sources of errors in the pulse.  We give a short account of the applicability 
of this scheme to a semiconductor device.

In section \ref{sec:qst} we briefly review the proposal for \qst\ given by
\cite{cir97} and give a new analytic form for a laser control pulse shape that implements
\qst.  In section \ref{sec:StochasticNoise} we formulate a master equation for the 
transfer subject to stochastic errors in the control pulse, and use the master 
equation to compute the sensitivity of the transfer process to such errors.  In 
section \ref{sec:optimise} we present a procedure to optimise the pulse shape 
to reduce the sensitivity of the transfer to control pulse errors.  We then finish 
with some conclusions.

\section{Quantum State Transfer} \label{sec:qst}

\subsection{Review of proposal} \label{sec:review}
 
\begin{figure}
\includegraphics[width=240pt]{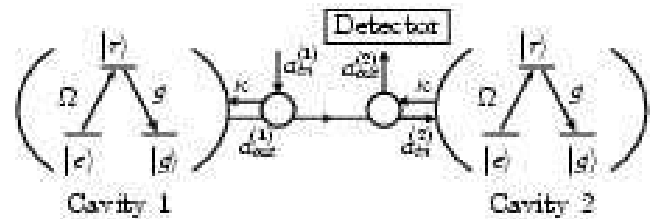}
\caption{\label{fig:DoubleCavity}
Schematic arrangement of cavity and three level system.  The small circles represent Faraday
isolators so the forward and reverse propagating paths are distinguishable.  Adapted from Cirac
\etal. \cite{cir97}.}
\end{figure}

In Cirac \etal.'s proposal for \qst, \cite{cir97}, two 3-level systems (atoms, in the original proposal) are located inside separate optical cavities, as illustrated in \fig{fig:DoubleCavity},
with the left atom starting in an arbitrary superposition of its lower levels,
$\ket{\psi}=c_g\ket{g}+c_e\ket{e}$, and the right atom in the state $\ket{g}$. 
The objective of \qst\ is then to transfer the state $\ket{\psi}$ from atom 1 to atom
2 by applying suitable laser pulses to each atom.  This may be expressed as
\begin{equation}
\ket{\psi}_1\ket{g}_2\otimes\ket{0}_1\ket{0}_2\rightarrow\ket{g}_1\ket{\psi}_2\otimes\ket{0}_1\ket{0}_2,
\end{equation}
where the subscripts indicate the atom/cavity system.  The term $\ket{0}_j$ is
cavity $j$'s photon state in the number basis, and is the vacuum
state at the start and end of the process.

In the case that the detuning between the laser frequency and the atomic
transition from $\ket{g}$ to $\ket{e}$, $\omega_L-\omega_0$, is larger than all other
frequencies in the problem, one can write a Hamiltonian for each atom-cavity system,
labeled $j=1, 2$, from which the upper level $\ket{r}$ is adiabatically eliminated,
\begin{equation}
\hat H_j(t)/\hbar=\frac{\Omega_j(t)^2}{4(\omega_L-\omega_0)}\ket{e}_{j j}\bra{e}-i g_j(t)(e^{i
\phi_j(t)}\ket{e}_{j j}\bra{g}\ahat{j} -\mathrm{H.c}).\label{eq:cavham}
\end{equation}
Here, $\Omega_j(t)=2 \boldsymbol{\mu}_{er}\cdot\boldsymbol{\varepsilon}E_j(t)/\hbar$
is the classical Rabi frequency of the atom-laser system and $g(t)=g
\Omega(t)/(\omega_L-\omega_0)$ is the effective coupling strength between state
$\ket{g}$ and $\ket{e}$, and is proportional to $E_j(t)$, the amplitude of the laser
electric field, which we will use to control the transfer.  We have also made the
following definitions: $\ahat{j}$ is cavity $j$'s photon creation operator, $g$ is the cavity-atom coupling strength, $\boldsymbol{\mu}_{er}$ is
the dipole matrix element between $\ket{e}$ and $\ket{r}$, $\boldsymbol{\varepsilon}$
is the laser polarisation, and $\phi_j(t)$ is the phase of the laser.  In deriving
this equation, there is an assumption that the laser frequency is detuned from
the cavity mode , $\omega_c$, nearest the transition, by an amount
{$\omega_l-\omega_c=g^2/(\omega_L-\omega_0)$}, which is the ac-Stark shift of
the state $\ket{g}_j$ in the electric field of the laser.

Following the original proposal by Cirac \etal. \cite{cir97}, the transfer is made to be unidirectional and is
constrained by the requirement that the photodetector, indicated in
\fig{fig:DoubleCavity}, never registers a photon.  The entire system is then an open
quantum system, and is described equivalently by a quantum master equation, or a
quantum trajectory punctuated by quantum jumps \cite{gar00}.  Since it will become
important for the later development of this paper, we summarise now some important
points from Cirac \etal., who adopt the quantum trajectories approach
to the problem.  The non-Hermitian effective Hamiltonian that describes the evolution
of the open system in between jumps is $ \hat H_{\mathrm{eff}}=\hat H_1+\hat H_2-i \kappa(\ahat{1}^\dagger\ahat{1}+\ahat{2}^\dagger \ahat{2}+2 \ahat{2}^\dagger \ahat{1})$, where $\kappa$ is the cavity leakage rate and the jump operator is $\hat c=\ahat{1}+\ahat{2}$.  In the language of quantum trajectories, the system state
vector evolves under the Schr\"odinger equation with the effective Hamiltonian
$H_{\mathrm{eff}}$.  This Hamiltonian does not increase the total excitation number
of the system, so we write the state vector as
\statevector
The initial conditions and final configuration (for a successful transfer) are 
\begin{equation}
\alpha_1(-\infty)=1, \phi_1(-\infty)=0\textrm{ and
}\alpha_2(+\infty)=1,\phi_2(+\infty)=0.\label{eq:ic}
\end{equation}
we refer to the condition at $t=+\infty$ as the zero-jump condition, which holds as 
long as as $\hat c\ket{\psi(t)}=0$ for all $t$, and this requires that
\begin{equation}
\beta_s(t)=0.\label{eq:zerojump}
\end{equation}
If we choose the phase of the laser to vary with time according to
$\dot\phi_j(t)=\Omega_j(t)^2/(4(\omega_L-\omega_0))$ then all $\phi_{j}$ dependence
factorises out of the Schr\"odinger, resulting in the linearly independent evolution equations
\begin{subequations}
\begin{eqnarray}
\dot\alpha_1(t)&=&g_1(t)\beta_a(t)/\sqrt{2},\label{eq:a1dot}\\
\dot\alpha_2(t)&=&-g_2(t)\beta_a(t)/\sqrt{2},\label{eq:a2dot}\\
\dot\beta_s(t)&=&g_1(t)\alpha_1(t)/\sqrt{2}+g_2(t)\alpha_2(t)/\sqrt{2}+\kappa\beta_a(t).\label{eq:bsdot}
\end{eqnarray} \label{eq:quanttraj}
\end{subequations}
Finally, the normalisation condition for the state vector requires that
\begin{equation}
|\alpha_1(t)|^2+|\alpha_2(t)|^2+|\beta_s(t)|^2+|\beta_a(t)|^2=1.\label{eq:norm}
\end{equation}
We could also write an evolution equation for $\beta_{a}$, however this is the same 
equation as obtained by differentiating \eqn{eq:norm} with respect to time, and then 
substituting \eqns{eq:a1dot} and (\ref{eq:a2dot}), so is not linearly independent.

There are thus only five independent equations, \eqns{eq:zerojump} through
(\ref{eq:norm}), but six unknowns, $\alpha_1, \alpha_2, \beta_s, \beta_a, g_1$ and
$g_2$, and so there is no unique solution that satisfies the requirements of \qst. 
There is a notable symmetry in the evolution equations and the initial and final
states which is that if the cavity labels $j=1,2$ are exchanged and time is reversed,
i.e. $1\leftrightarrow 2$ and $t\rightarrow-t$, then they are recovered, so long as
$g_2(t)=g_1(-t)$, which is referred to as the symmetric pulse condition.  Under this
condition, Cirac \etal.\ \cite{cir97} present a method for generating pulse shapes that satisfy
the requirements of \qst.  This procedure is as follows:
\begin{enumerate}
\item \noindent Select $g_1(t)$ for $t\geq0$.  \item Solve \eqn{eq:a1dot} and a reduced form
of \eqn{eq:bsdot}: {$\dot\beta_a(t)=-g_1(t)\beta_a(t)/\sqrt{2}-\kappa
g_1(t)\alpha_1(t)$}, subject to the initial conditions
{$\alpha_1(0)=\kappa/\sqrt{2(g_1(0)^2+\kappa^2)}$} and
{$\beta_a(0)=-\sqrt{1-2\alpha_1(0)}$}.  
\item Define \linebreak[3]
{$g_1(t)=-(\sqrt{2}\kappa\beta_a(|t|)+g_1(|t|)\alpha_1(|t|))/\alpha_2(|t|)$} for {$t<0$}.
\end{enumerate}
Then $g_1(t)$ is defined at all times, and as long as $g_1(+\infty)>0$, this pulse shape
will satisfy the requirements of \qst.   Cirac \etal.\ \cite{cir97}  present a particular solution to the
equations, which they show accomplishes the transfer successfully.

We make the point that if $g_{1}(t)\rightarrow0$ as $t\rightarrow+\infty$ the method
doesn't guarantee a successful transfer pulse shape, but it does allow us to evaluate
the success of a transfer based only on knowledge of $g_{1}(t)$ for $t\geq0$, subject
to the presumption that $g_{1}(t): t<0$ is determined from $g_{1}(t): t\geq0$,
according to the above procedure.

\subsection{Alternate solution method}

Using \eqns{eq:a1dot} and (\ref{eq:a2dot}) to replace $g_1(t)$ and $g_2(t)$ in
\eqn{eq:bsdot}, along with the zero jump condition, \eqn{eq:zerojump}, as well as
the normalisation condition \eqn{eq:norm} results in the following reduced non-linear equation
for $\alpha_1(t)$ and $\alpha_2(t)$,
\begin{equation}
\frac{d}{dt}\left(\frac{\alpha_1(t)^2-\alpha_2(t)^2}{2}\right)+\kappa(1-\alpha_1(t)^2-\alpha_2(t)^2)=0.
\label{eq:redevol}
\end{equation}
Thus, another approach to finding a suitable pulse shape is to invent another
equation depending on $\alpha_1(t)$ and $\alpha_2(t)$ (i.e.
$f(\alpha_1(t),\alpha_2(t))=0$) consistent with \eqns{eq:ic} and (\ref{eq:norm})
and then solve this equation and \eqn{eq:redevol} to find $\alpha_1(t)$ and $\alpha_2(t)$, from which we
can calculate the other functions.

We can motivate an example of this method by making $g_1(t)=g_2(t)$, so that pulses
on each atom are the same.  We then divide \eqn{eq:a1dot} by \eqn{eq:a2dot}, to find
{$\dot \alpha_1(t)=-\dot \alpha_2(t)$}, and taking account of the initial condition,
\eqn{eq:ic}, we find that $\alpha_1(t)+\alpha_2(t)=1$, which is of the form
$f(\alpha_1(t),\alpha_2(t))=0$, and is consistent with \eqns{eq:ic} and
(\ref{eq:norm}).  Thus we can use this as a second equation, along with
\eqn{eq:redevol} to solve for $\alpha_1(t)$ and $\alpha_2(t)$.  By substituting for
$\alpha_{2}(t)$ in \eqn{eq:redevol}, we find that $\alpha_1(t)$ satisfies
\begin{equation}
\dot \alpha_1(t)+2\kappa(\alpha_1(t)-\alpha_1(t)^2)=0.\label{eq:a1eqn}
\end{equation}
There are two solutions to this equation, {$\alpha_1(t)=(1\pm e^{2\kappa t})^{-1}$},
however one is unphysical since it diverges at $t=0$ (violating the normalisation
condition).  The physical solution is $\alpha_1(t)=(1- \tanh(\kappa t))/2$,
and from this we can determine all the other quantities to be
\begin{subequations}
{\begin{eqnarray}
\alpha_2(t)&=&\alpha_1(-t)=(1+\mathrm{tanh}(\kappa t))/2,\label{eq:a1sol}\\
\beta_a(t)&=&-\mathrm{sech}(\kappa t))/\sqrt{2},\label{eq:basol}\\
g_1(t)&=&g_2(t)=\kappa\, \mathrm{sech}(\kappa t)\vphantom{\frac{\mathrm{sech}(\kappa
t)}{\sqrt{2}}}.\label{eq:g1sol}
\end{eqnarray}}\label{eq:tsoln}
\end{subequations}
\begin{figure}
\includegraphics[width=240pt]{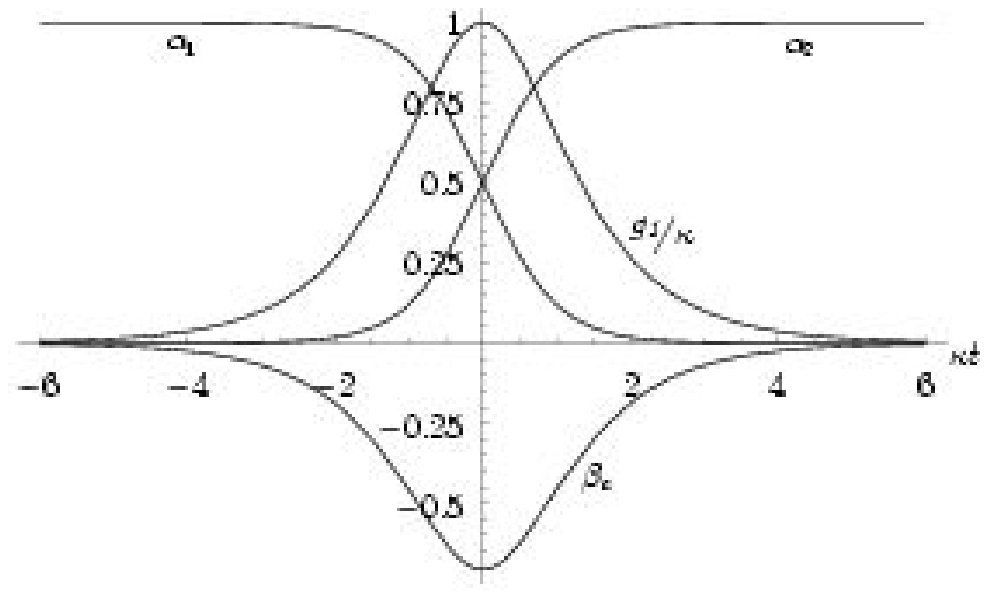}
\caption{\label{fig:SechPulse}
Plot of the quantities given by \eqns{eq:a1sol} to (\ref{eq:g1sol}).}
\end{figure}
Notably, this solution has the property that $g_1(t)\rightarrow0$ as
$t\rightarrow\infty$, and it also satisfies the symmetric pulse condition,
$g_2(t)=g_1(-t)$, so is a member of the class of solutions generated by the method of
Cirac \etal.\ \cite{cir97}.  We now have two solutions out of an infinite range of possible ones
(i.e. the solution presented in \cite{cir97}, and the one above), which we will
compare in later sections.

\section{Stochastic Noise in Laser Pulse}\label{sec:StochasticNoise}

Having presented an alternate pulse shape to accomplish \qst, we now use the two
solutions, that in Cirac \etal.\ \cite{cir97}  and in this paper, as test cases for 
evaluating the effect of stochastic noise in the laser pulses on the fidelity of the
transfer.  We reformulate the problem in terms of a
quantum master equation, as opposed to that of quantum trajectories \cite{enk97a, enk97, cir97}, since we will
eventually take averages over an ensemble of noise realisations, which must be done with
respect to the density matrix rather than the state vector.

\subsection{Master Equation}

\begin{figure}
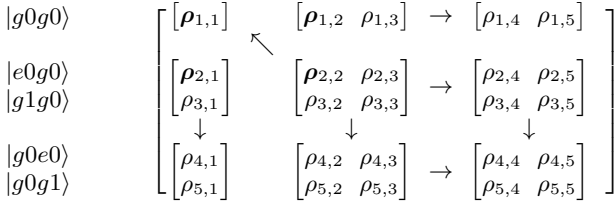

\begin{displaymath}
	\begin{matrix}
		\ket{g0g0}\\ \\ \ket{e0g0}\\ \ket{g1g0}\\ \\ \ket{g0e0}\\\ket{g0g1}
	\end{matrix}
	\hspace{1cm}
	\begin{bmatrix}
		\begin{bmatrix}
			\boldsymbol\rho_{1,1}\end{bmatrix}&&\begin{bmatrix}\boldsymbol\rho_{1,2}&\rho_{1,3}\end{bmatrix}&\rightarrow
			&\begin{bmatrix}\rho_{1,4}&\rho_{1,5}\end{bmatrix}\\
			&\nwarrow&&&\\
			\begin{bmatrix}
				\boldsymbol\rho_{2,1}\\\rho_{3,1}\end{bmatrix}&&\begin{bmatrix}\boldsymbol\rho_{2,2}&\rho_{2,3}\\\rho_{3,2}&\rho_{3,3}\end{bmatrix}&\rightarrow
				&\begin{bmatrix}\rho_{2,4}&\rho_{2,5}\\\rho_{3,4}&\rho_{3,5}\end{bmatrix}\\
				\downarrow && \downarrow && \downarrow &\\
				\begin{bmatrix}\rho_{4, 1}\\ 
					\rho_{5,1}\end{bmatrix}&&\begin{bmatrix}\rho_{4,2}&\rho_{4,3}\\\rho_{5,2}&\rho_{5,3}\end{bmatrix}&\rightarrow
					&\begin{bmatrix}\rho_{4,4}&\rho_{4,5}\\\rho_{5,4}&\rho_{5,5}\end{bmatrix}
	\end{bmatrix}
\end{displaymath}
\caption{Structure of relationships between density matrix elements arising from
\eqn{eq:master}, with respect to the basis $\mathcal{B}$, which is indicated on the
left.  Elements within each sub-matrix are mutually coupled.  Arrows indicate the
direction of coupling between the sub-matrices, so for instance $\dot \rho_{5,5}$
depends on $\rho_{3,5}$, but not vice-versa, i.e. derivatives are placed at the arrow
head, for example the equation for $\dot{\rho}_{{5,5}}$ is {$\dot{
\rho}_{5,5}={g_2}(t)\left( \rho_{4,5} + \rho_{5,4} \right) - 2\kappa\left( \rho_{3,5}
+ \rho_{5,3} + \rho_{5,5} \right) $}.  The arrows therefore indicate the flow of
information, from the states with a single excitation in the left cavity to single
excitation in the right cavity, but not in reverse, as determined by the one-way
requirement used to formulate the equations.  The bold elements are those that may be
non-zero at $t=-\infty$, and are given by
$\ket{\psi}\bra{\psi}$.}\label{fig:coupling}
\end{figure}

Since the process of information transfer is one-way, we adopt the cascaded quantum
system approach, described in the text by Gardiner and Zoller \cite{gar00}, to 
derive a master equation.  The
quantum master equation for the cascaded system here is given by
\begin{subequations}
	\begin{gather}
		\dot \rho=-\frac{i}{\hbar}\left[\hat H_1(t)+\hat H_2(t),\rho\right]+\kappa
		\mathscr{L}\{\rho\},\label{eq:master}\\
		\begin{split}
			\mathscr{L}\{\rho\}=&2 \ahat{1} \rho \ahat{1}^\dagger-\ahat{1} \ahat{1}^\dagger \rho-\rho\ahat{1}
			\ahat{1}^\dagger\\
			&+2 \ahat{2} \rho \ahat{2}^\dagger-\ahat{2} \ahat{2}^\dagger\rho-\rho\ahat{2}\ahat{2}^\dagger\\
			&-2\{[\ahat{2}^\dagger,\ahat{1}\rho]+[\rho\ahat{1}^\dagger,\ahat{2}]\},\label{eq:lindblad}
		\end{split}
	\end{gather}
\end{subequations}
where $\rho$ is the system density matrix.  Subsequently, we leave out explicit time
dependence notation from the Hamiltonians, $H_{1}$ and $H_{2}$, unless required for clarity.  In lieu of
writing out all 25 coupled, first-order differential equations arising from the master equation,
we give a schematic representation of the coupling between elements of the density
matrix in \fig{fig:coupling}, where arrows indicate the direction of the coupling. 
From this schematic representation we can see how the equations ensure that information does
flow unidirectionally from the left cavity to the right.  The initial condition for
this master equation is simply
\begin{equation}
	\rho(-\infty)=(\ket{\psi}_1\ket{g}_2\ket{00})({}_{1}\bra{\psi}_2\bra{g}\bra{00}).
\label{eq:masteric}
\end{equation}
For future reference, if a matrix representation of an operator is given, it will be
expressed with respect to the basis
{$\mathcal{B}=\{\ket{g0g0},\ket{e0g0},\ket{g1g0},\ket{g0e0},\ket{g0g1}\}$},
where the state vector notation represents the state of the left atom, left
photon number, right atom and right photon number in order.  The solutions we are interested in
will be those for which \qst\ has been successfully implemented, so
\begin{equation}
	\rho(+\infty)=(\ket{g}_1\ket{\psi}_2\ket{00})({}_{1}\bra{g}_2\bra{\psi}\bra{00}).
\label{eq:masterfc}
\end{equation}
We will use the notation $\rho_0(t)$ to indicate any solution to the noiseless master
equation, \eqn{eq:master}, which satisfies both \eqns{eq:masteric} and 
(\ref{eq:masterfc}).  For an arbitrary time $t$, $\rho_{0}(t)$ will depend on the 
pulse shape, nevertheless all permissible solutions will agree at $t=\pm\infty$.  Of 
course, for those pulses that do satisfy the requirement of \qst, the master equation 
will give the same result as solving the quantum trajectory equations, 
\eqn{eq:quanttraj}.

\subsection{Adding Stochastic Noise}

We add the effect of stochastic noise in the pulse shapes to the dynamics of the
problem, and so we follow a similar development to that presented in recent work
\cite{bud01,wel01}.  We assume that for each realisation of a transfer attempt, the
actual classical laser pulse, $\tilde g_j(t)$, applied to atom $j$ is perturbed from
the desired shape according to {$\tilde g_j(t)=g_j(t)+\xi_j(t)\Delta g_j(t)$},
where $\xi_j(t)$ is a stochastic function of time, and $\Delta g_j(t)$ is some
function which will depend on the origin of the stochastic error (i.e. noise), and is
considered to be small.  For instance, if the noise is associated with fluctuations
in the amplitude of the pulse, then $\Delta g_j(t)=g_j(t)$, so that {$\tilde
g_j(t)=(1+\xi_j(t))g_j(t)$}, and if it is associated with errors in timing between
the pulses applied to the two cavities then $\Delta g_j(t)=\dot{g_j}(t)$, both of
which we consider later.  This modifies the atom-cavity Hamiltonian from that with an
ideal pulse, which to first order in $\xi_j(t)$ is
\begin{equation}
\hat H_j(t)\rightarrow \hat H_j(t)+\xi_j(t)\hat h_j(\Delta g_j(t)).\label{eq:hamsub}
\end{equation}
The operator $\hat h_j$ is Hermitian and depends on $\Delta g_j(t)$.

With the modified Hamiltonian given in \eqn{eq:hamsub}, the master equation reads
\begin{equation}
\dot \rho=-\frac{i}{\hbar}\left[\hat H_1+\xi_1(t)\hat h_1(t)+\hat
H_2+\xi_2(t)\hat h_2(t),\rho\right]+\kappa\mathscr{L}\{\rho\},
\end{equation}
which we formally integrate on both sides to find $\rho(t)$ and substitute back into
the right-hand side (RHS) of the equation, to generate the second order term in the
Dyson series.  We then take averages over an ensemble of noise realisations, and
assuming the one-time correlation terms have zero mean, i.e.\ $\braket{\xi_j(t)}=0,$ we arrive
at the result
\begin{widetext}
\begin{equation}
	\begin{split}
		\braket{\dot\rho(t)}=&-i/\hbar\commute{H_1+H_2}{\braket{\rho(t)}}+\kappa\mathscr{L}\{\braket{\rho(t)}\}\\
		&+(i/\hbar)^2\left(
		\commute{\hat h_1}{\int_{-\infty}^t dt'\braket{\commute{\xi_1(t)\xi_1(t')\hat h_1(t')}{\rho(t')}}}+
		\commute{\hat h_2}{\int_{-\infty}^t dt'\braket{\commute{\xi_2(t)\xi_2(t')\hat h_2(t')}{\rho(t')}}}
		\right.\\ 
		&\hphantom{+(i/\hbar)^2}+\left.
		\commute{\hat h_1}{\int_{-\infty}^t dt'\braket{\commute{\xi_1(t)\xi_2(t')\hat h_2(t')}{\rho(t')}}}+
		\commute{\hat h_2}{\int_{-\infty}^t dt'\braket{\commute{\xi_2(t)\xi_1(t')\hat h_1(t')}{\rho(t')}}}\right).
		\label{eq:stochmaster}
	\end{split}
\end{equation}
\end{widetext}
We write the averaged solution to the noisy master equation, \eqn{eq:stochmaster}, as a
correction to $\rho_{0}(t)$, the density matrix in the absence of noise,
\begin{equation}
\braket{\rho(t)}=\rho_0(t)+\epsilon_1 \dr_1(t)+\epsilon_2 \dr_2(t),\label{eq:meanrho}
\end{equation}
where $\epsilon_j$ is a measure of the (small) variance of the noise on laser pulse 
$j$, and will appear
as a coefficient in the two-time correlation function $\braket{\xi_j(t)\xi_j(t')}$. 
In the case that the stochastic terms on each pulse are uncorrelated with one
another, $\braket{\xi_1(t)\xi_2(t')}=0$, which we will hereafter assume, we can
substitute \eqn{eq:meanrho} into \eqn{eq:stochmaster} and use \eqn{eq:master} to
derive two independent master equations for the first order (in $\epsilon_j$)
correction terms $\dr_1(t)$ and $\dr_2(t)$,
\begin{equation}
	\begin{split}
		\dot{\dr_j}=&-i/\hbar\commute{\hat H_1+\hat H_2}{\dr_j}+\kappa\mathscr{L}\{\dr_j\}\\
		&+(i/\hbar)^2 \commute{\hat h_j}{\int_{-\infty}^t
		dt'\commute{\frac{\braket{\xi_j(t)\xi_j(t')}}{\epsilon_j}\hat h_j(t')}{\rho_0(t')}}.
	\end{split}
\label{eq:mastercorr}
\end{equation}
The quantity $\epsilon_{j}\dr_j(t)$ is therefore interpreted as a correction to  the noiseless density matrix due to stochastic errors in pulse $j$.
We note that the homogeneous part of the above equation is the same
as the original noiseless master equation, \eqn{eq:master}, but it now has an
inhomogeneous driving term that depends on the solution to the noiseless equation,
$\rho_0(t)$.  We are interested in errors in the density matrix associated with
stochastic errors in the pulse shape, so we assume that the initial state of the
density matrix can be prepared exactly, i.e $\dr_{j}(-\infty)=0$.  The solution to
equation \eqn{eq:master} as $t\rightarrow\infty$ is determined by \eqn{eq:masterfc},
but the solution to \eqn{eq:mastercorr} is not constrained and may depend on both $g_{j}(t)$
and $\Delta g_{j}(t)$.

In numerical investigations of the error, we assume that the initial state of
the left atom is $\ket{\psi}=\ket{e}$, so that the only non-zero matrix element is
$\bra{e0g0}\rho_{0}(-\infty)\ket{e0g0}=1$.  This assumption is not restrictive, since
as may be seen from the form of the coupling of the evolution equations in
\fig{fig:coupling}, the only initial condition that influences the matrix elements
$\rho_{i,j}(t), i,j\geq 2$ is that on the element {$\rho_{2,2}(-\infty)=
\bra{e0g0}\rho_{0}(-\infty)\ket{e0g0}=1$}, and by the linearity of the equations we
obtain the solution (up to a scaling factor) for any non-zero initial condition.

At the end of a noisless transfer, the only non-zero element will be
$\bra{g0e0}\rho_{0}(+\infty)\ket{g0e0}=1$, but at that time the correction to the
density matrix may be non-zero, and the corresponding element
{$\eta_{j}(t)=\bra{g0e0}\dr_{j}(t)\ket{g0e0}$} (which depends on $\Delta
g_{j}(t)$) we call the noise sensitivity.  This is motivated by \eqn{eq:meanrho},
since errors in pulse $j$ will perturb the final result away from unity
\begin{equation}
	\bra{g0e0}\rho(+\infty)\ket{g0e0}
	=1+\epsilon_{j}\eta_{j}(+\infty).
	\label{eq:noisesensitivity}
\end{equation}
We can conclude from this that $\eta_{j}(+\infty)\leq0$, since $\epsilon_{j}\geq 0$ and 
$\tr(\rho)=1$.  The relation above makes evident the fact that the stochastic (classical) noise 
is a source of decoherence for the transfer, since the probability of finding the 
system in the state $\bra{g0e0}\rho(+\infty)\ket{g0e0}$ is less than unity.

\subsection{White Amplitude Error}

We first apply the above result to the case where the stochastic errors arrive
from unwanted fluctuations in the amplitude of the pulse, as mentioned earlier.  In
principle this could arise from intrinsic laser amplitude noise, although this is
typically negligibly small.  More likely sources of error are in the details of the
pulse shaping and might arise from phase and amplitude errors of the
modes selected to generate the pulse.  In any case, we do not consider these issues
here, but instead assume that the error has the property that $\Delta g_j(t)=g_j(t)$,
so that
\begin{equation}
	\tilde g_j(t)=(1+\xi_j(t))g_j(t).\label{eq:ampnoise}
\end{equation}
We will also assume that the stochastic terms are $\delta$-correlated,
\begin{equation}
	\braket{\xi_j(t)\xi_{j}(t')}=\epsilon_{j}\delta(t-t').
	\label{eq:deltacorrel}
\end{equation}
With these assumptions the master equation for the corrections to the density matrix
reduce to
\begin{equation}
\dot{\dr_j}=-\frac{i}{\hbar}\commute{\hat H_1+\hat H_2}{\dr_j}+\mathscr{L}\{\dr_j\}
-\frac{1}{2 \hbar^2} \commute{\hat h_j}{\commute{\hat h_j}{\rho_0}},
	\label{eq:ampcorrmat}
\end{equation}
where the factor of $1/2$ in front of the double commutator arises from integrating only
half of the Dirac-$\delta$ function, and \eqn{eq:ampcorrmat} is of Lindblad form.

\begin{figure}
	\centering
	\includegraphics[width=240pt]{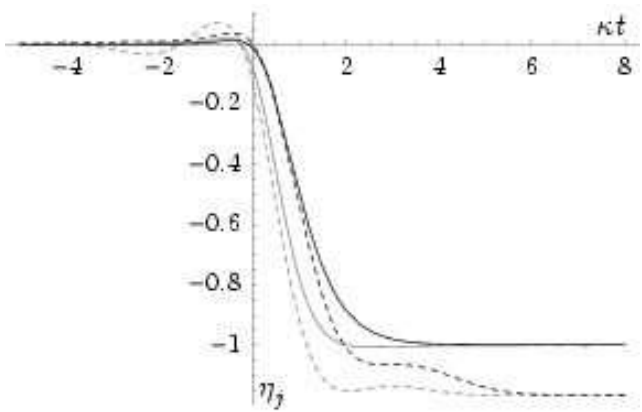}
	\caption{The evolution of the noise sensitivities, $\eta_{j}$, for amplitude
	errors.  The upper curves (solid) correspond to the pulse shapes given in this paper, in
	\eqns{eq:tsoln}, whilst the lower curves (dotted) are for the pulse shapes presented
	elsewhere \cite{cir97}.  The darker traces correspond to the solution for $\eta_{1}(t)$
	lighter traces are for $\eta_{2}(t)$.}
	\label{fig:ampcorr}
\end{figure}
We have solved the equations numerically (using the numerical differential equation
solver provided with Mathematica 4.0) for two different pulse shapes, one given in
this paper, \eqns{eq:tsoln}, and the other taken from the proposal for \qst\
by Cirac \etal.\ \cite{cir97}.  The results for the noise sensitivities are plotted in
\fig{fig:ampcorr}.  For each pulse shape the noise sensitivities $\eta_{1}(t)$ and
$\eta_{2}(t)$ appear to converge to the same value as $t\rightarrow\infty$, however
they do differ by a small but definite amount between pulses shapes, being $-1$ (to
the numerical precision of the equation solver) and approximately $-1.17$.

\subsection{White Timing Error}

Using the same methods, we can consider the effect of timing errors between the
pulses applied to the cavities.  In this case it makes sense to define only one noise 
sensitivity, $\eta$, since we may assume that the pulse on the first atom defines the origin
of time, and thus it will be errors in the timing of the second pulse relative to the
first that will contribute to state transfer errors.

For timing errors, the noisy pulse shape is given by
\begin{eqnarray}
	\tilde g_{2}(t)&=&g_{2}(t+\xi(t)),\nonumber\\
	&\approx&g_{2}(t)+\xi(t)g_{2}'(t),
	\label{eq:timenoise}
\end{eqnarray}
where the second line follows by expanding to first order in the quantity $\xi(t)$. 
It may be seen from this form that $\Delta g_{2}(t)=g_{2}'(t)$, and for the pusposes
of illustration, we will again take the errors to be $\delta$-correlated,
$\braket{\xi(t)\xi(t')}=\epsilon\delta(t-t')$.

\begin{figure}
	\centering
	\includegraphics[width=240pt]{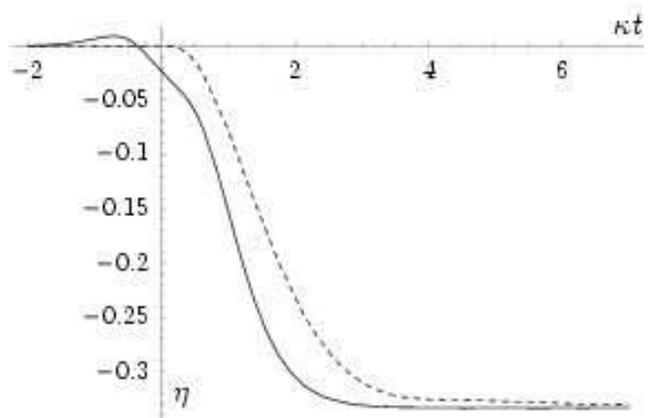}
	\caption{The evolution of the noise sensitivities, $\eta$, for timing errors. 
	The lower curve corresponds to the pulse shapes given in this paper, in
	\eqns{eq:tsoln}, whilst the upper curve is for the pulse shapes presented
	elsewhere \cite{cir97}.}
	\label{fig:timecorr}
\end{figure}

The result of the numerical solutions for the corrections to the density matrix are 
shown in \fig{fig:timecorr}.  In this case the pulse shape presented in this paper is 
very slightly more susceptible to timing errors than the pulse shape given by Cirac 
\etal. \cite{cir97}.

\section{Optimising against noise}\label{sec:optimise}

From the two examples presented above, we can see that different pulse shapes are 
more or less sensitive to errors in their generation, depending on the nature of the 
error.  We will now examine how this variation may be used to optimise the pulse 
shape against sources of error.  

The utility of this is in recognising that stochastic noise in control signals
contributes to decoherence, which has been noted previously \cite{wel01}.  Whilst in
that work, the decoherence attributable to stochastic noise was found to be much less
than that due to other effects (environment), the calculation was based on a
nominally constant control signal with stochastic noise superposed.  The situation we
consider here, though in a different system, is slightly more general in that the
Hamiltonian is explicitly time dependent.  Thus the errors in the control pulse will
not be simply the intrinsic (e.g.\ thermal) noise associated with the power source,
but also dynamic errors associate with signal generation, e.g. due to digitisation of
a nominally continuous pulse, or errors in timing.

\subsection{Optimisation Strategy}

In principle, the problem of optimising the pulse shape against noise may be
understood in terms of constrained optimisation.  We wish to choose a pulse shape
that minimises the noise sensitivity, $\eta_{j}(+\infty)$, subject to the constraint that it
satisfies the condition of pulse transfer, i.e. $\alpha_{2}(+\infty)=1$.  
Conceptually, this would provide us with another equation to supplement \eqns{eq:quanttraj} 
which then makes the number of equations and unknowns equal, thereby specifying 
the pulse shape uniquely.  In practise, we could formulate this problem in terms of Lagrange 
multipliers \cite{lagmult}, and derive the following constrained optimisation 
equations
\begin{eqnarray}
	\frac{\vd \eta_{j}(+\infty)}{\vd g_{1}(\tau)}+
	\lambda \frac{\vd \alpha_{2}(+\infty)}{\vd g_{1}(\tau)}&=&0,\label{eq:lagmult}\\
	\alpha_{2}(+\infty)=1,\label{eq:lagconst}
\end{eqnarray}
where the operator ${\vd}/{\vd g_{1}(\tau)}$ represent the functional derivative with
respect to the function $g_{1}(\tau)$, and $\lambda$ is an unknown Lagrange
multiplier.  We have introduced two more equations, and one more unknown, to bring
the total number of each to seven.  Since we have chosen $g_{2}(\tau)=g_{1}(-\tau)$
we do not consider functional derivatives with respect to $g_{2}(t)$.  Although
we don't prove it formally, we find that $\eta_{1}(+\infty)=\eta_{2}(+\infty)$ in all numerical
calulations we have performed, so we only consider variational derivatives with
respect to $\eta_{1}(+\infty)$.

In fact, computing either of the above functional derivatives requires the solution
of a set of differential equations that are of essentially the same form as
\eqn{eq:master}, since they are formulated by taking variational derivatives of the
master equation with respect to $g_{1}(\tau)$, which must then be integrated.  Since
we do not have an analytic solution to the master equations for an arbitrary pulse
shape, it is difficult to see how practically to make use of these equations, except
for very small scale discretisations.

Another approach is to use numerical substitution.  Here, we discretise the pulse
shape, defining $g_{j}=g_{1}(t_{j})$ for several discrete times $t_{j}=t_{0}, \ldots,
t_{n-1}$ (setting $g_{n}=0$), then interpolating between points.  We allow all but
$g_{0}$ to be variational parameters, and $g_{0}$ is chosen depending on $g_{1},
\ldots, g_{n-1}$ to satisfy the constraint, $\alpha_{2}(+\infty)=1$, which is the
numerical equivalent of a substitution for $g_{0}$ in terms of all other
variational parameters.  We then have $n-1$ variational parameters, $g_{1}, \ldots,
g_{n-1}$, available for unconstrained optimisation of $\eta_{1}(+\infty)$, for which
we may use an unconstrained optimisation scheme.  In the details of this
method, the discretisation points are chosen to be in the domain $t\geq0$, and
$g_{0}$ is determined from $g_{1}, \ldots, g_{n-1}$ by use of the method outlined by 
Cirac \etal.\ \cite{cir97}  (and at the end of section \ref{sec:review} in this paper) which 
determines the efficacy of a given pulse shape in terms of its form for $t\geq 0$.

\subsection{Numerical Results}
\begin{figure*}
	\renewcommand{\subfigcapskip}{-10pt}
	\renewcommand{\subfigbottomskip}{0pt}
	\renewcommand{\subfigtopskip}{10pt}
	\subfigure[]{\includegraphics[width=200pt]{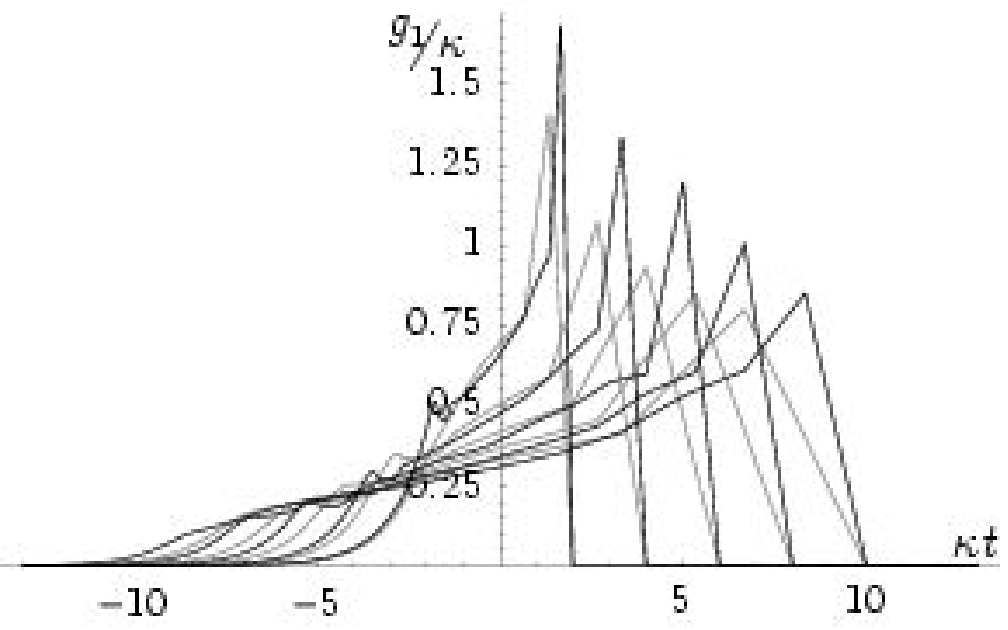}\label{fig:pulse1}}
		\hspace{1cm}
	\subfigure[]{\includegraphics[width=200pt]{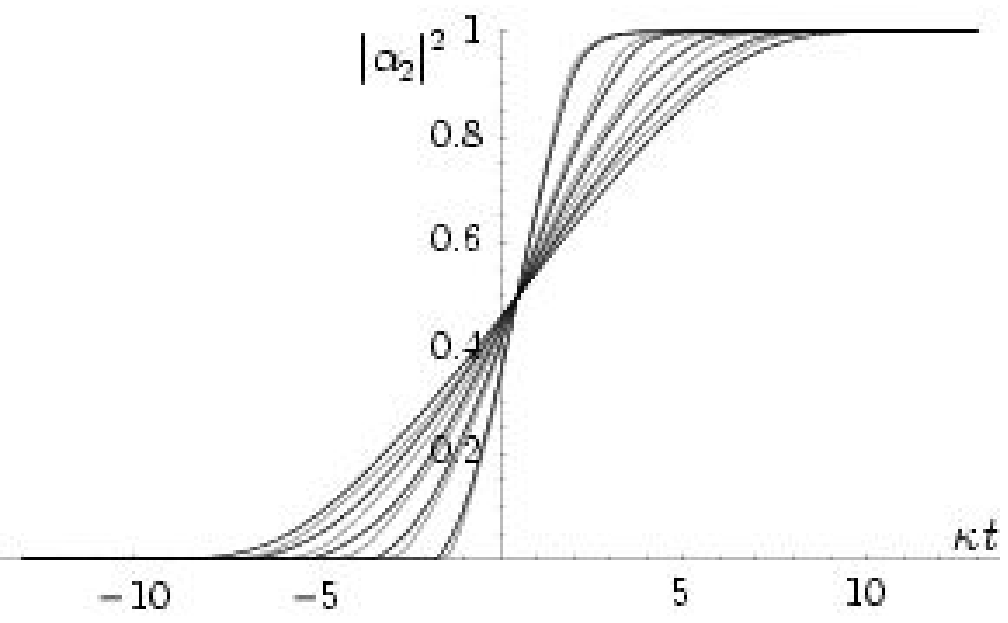}\label{fig:a21}}
	\subfigure[]{\includegraphics[width=200pt]{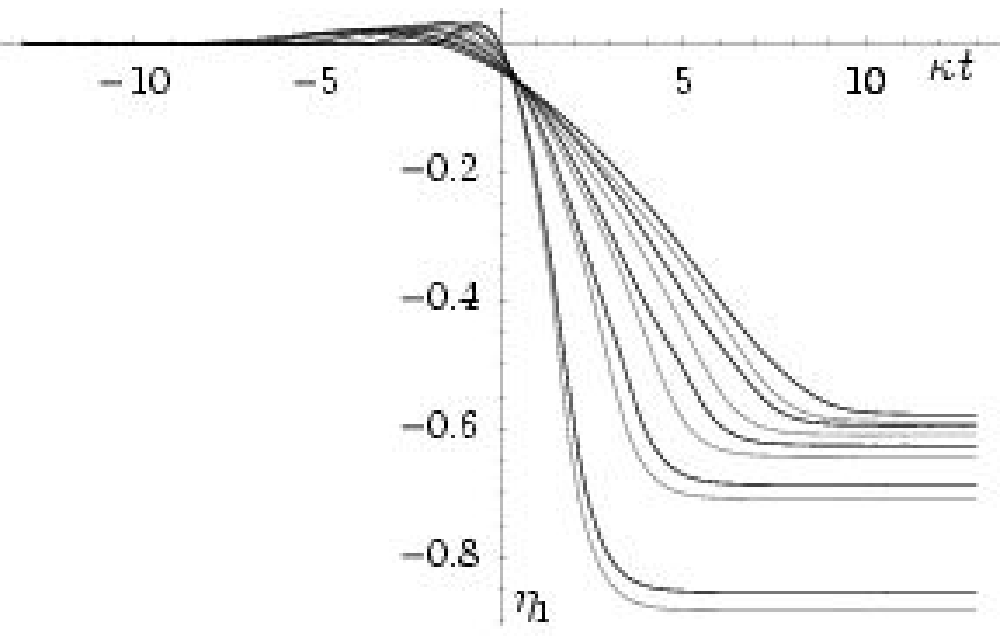}\label{fig:eta1}}
		\hspace{1cm}
	\subfigure[]{\includegraphics[width=200pt]{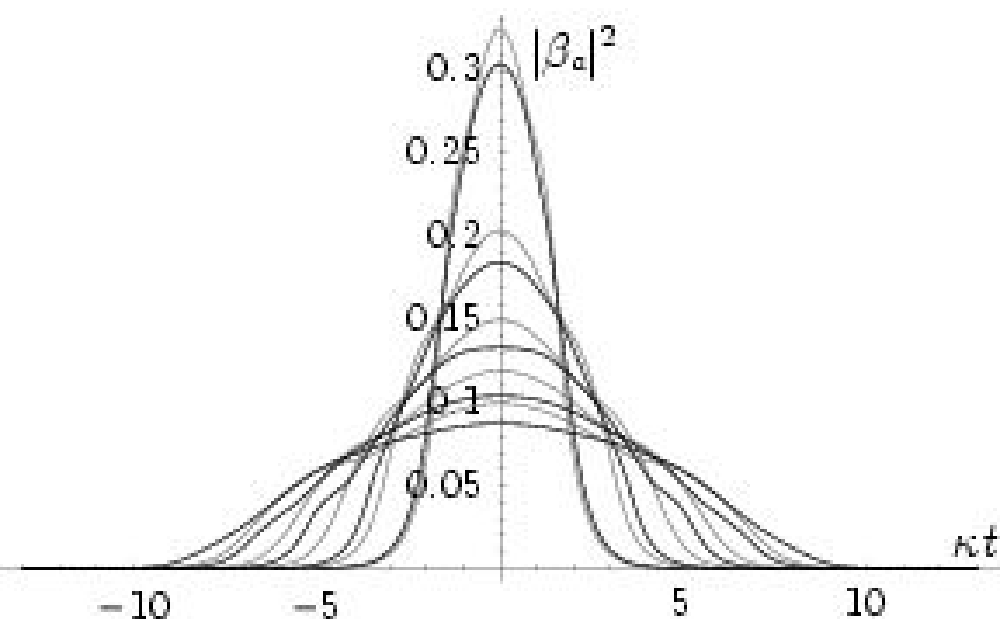}\label{fig:beta}}
	\subfigure[]{\includegraphics[width=200pt]{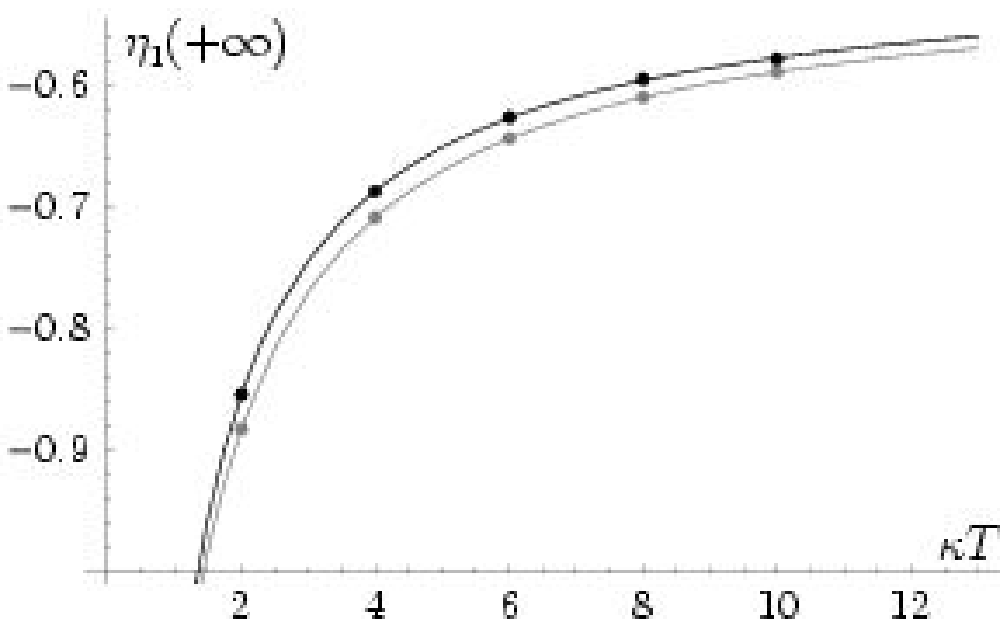}\label{fig:extrapolation}}
	\caption{(a) Optimised pulse shapes using three equally spaced points (light),
	and six equally spaced points (dark) for each of five different pulse widths. 
	(b) Plot of $\alpha_{2}(t)^{2}=\bra{e0g0}\rho_{0}(t)\ket{e0g0}$ for each of the
	pulses.  (c) Plot of $\eta_{1}(t)$ showing noise sensitivity for the same 
	pulses. 
	(d) $\beta_{a}(t)^{2}=2\bra{g1g0}\rho_{0}(t)\ket{g1g0}$ for each pulse.  
	(e) Showing $\eta_{1}(+\infty)$ versus $T$ (points), and a fit to the data
	(solid) using a hyperpbolic relation.
	In all figures, traces from steepest to shallowest correspond to narrowest to
	broadest pulses in (a), with light traces corresponding to three-point
	optimisations, and dark traces to six-point optimisations.}
	\label{fig:optimum}
\end{figure*}

A number of numerical optimisations are performed according to the numerical
substitution scheme outlined above.  In one set of optimisations, three points are
allowed vary, and one of them, at $t=0$, is used to satisfy the condition for pulse
transfer.  The locations of the other two points are equally spaced between $t=0$ and
a fourth location, $t=T$, where the function is zero, i.e. the two intermediate
points are located at $t=T/3$ and $t=2T/3$.  These intermediate points are varied to
optimise the pulse shape.  The pulse shapes that result from the three-point optimisation process
are presented as the lighter traces in \fig{fig:pulse1} for five different values of
$\kappa T=2,4,6,8$ and 10.

In the other set of optimisations, six points are allowed to vary, again with the 
one at $t=0$ being used to satisfy the condition for successful pulse transfer.  
The remaining five points are also equally spaced between $t=0$ and $t=T$, for the 
same values of $T$ as above, and the optimal pulse shapes for the six-point 
optimisations are shown as the darker traces in \fig{fig:pulse1}.

Figure \ref{fig:a21} demonstrates that the various pulses shown in \fig{fig:pulse1}
do each implement \qst\ successfully, since $|\alpha_{2}(t)|^{2}\rightarrow1$ as
$t\rightarrow\infty$ for all pulses.  The figure also shows that as the width of the pulse becomes broader,
proportional to $T$, the transfer process takes longer, so the slope of 
$|\alpha_{2}(t)|^{2}$ becomes shallower.

Figure \ref{fig:eta1} shows the noise sensitivity, $\eta_{1}(t)$, for each of the
pulses in \fig{fig:pulse1}.  Naturally, the value of $\eta_{1}(+\infty)$ gets closer
to zero (i.e. better) as the number of optimisation points is increased, and it also
gets better as the width of the pulse increases.  It may be compared with
\fig{fig:ampcorr} which shows the noise sensitivity for the two analytic solutions
discussed earlier and the optimisation process has clearly produced pulses that are
significantly less sensitive to noise in the laser pulses than either of the analytic
solutions discussed earlier in this paper.  The best pulse shape computed in the
optimisations performed here is that with $T=10$, and has the limit
$\eta_{1}\rightarrow-0.58$ as $t\rightarrow\infty$.  It appears that the optimisation
procedure selects solutions with lower amplitudes, and this occurs because
the form of noise we have assumed in this set of examples is proportional to the
amplitude of the control pulse, thus lowering the amplitude tends to lower the 
effect of the noise.

Figure \ref{fig:beta} shows that the photon wave-packet transmitted between the
cavities, which is proportional to $\beta_{a}(t)$, becomes broader.  It may be seen
from \eqns{eq:norm} and (\ref{eq:redevol}) that
$\int_{-\infty}^{+\infty}|\beta_{a}(t)|^{2}dt=1$, and the numerical solutions satisfy
this to around one part in $10^{5}$.

Figure \ref{fig:extrapolation} plots the limiting value of $\eta_{1}$ versus $T$, the
time after which $g_{1}$ is zero, for the three-point (light) and six-point (dark)
optimisations.  The solid curves are fitted using a function of the form
$a+\frac{b}{c+T}$, where $a,b$ and $c$ are fitting parameters.  In both cases, the
surprisingly close fit has $a=-0.5$ to better than one part in $10^{-3}$, suggesting
that the ultimate limit to the optimisation process is an infinitely wide pulse with
$\eta_{1}(+\infty)=-0.5$.  This indicates that the effect of the noise cannot be 
removed by allowing the pulse duration to become infinitely long.

\section{Discussion}

The results presented in \fig{fig:optimum} above show that there is a clear 
possibility of selecting pulse shapes that are more or less sensitive to fluctuations 
therein.  Figures \ref{fig:ampcorr} and \ref{fig:timecorr} show that 
pulses which perform better with respect to one source of error may perform worse 
with respect to another, so in a realistic implementation of the optimisations 
presented here, detailed knowledge of the sources of stochastic error in the pulses 
would be required.

The reduction in transfer error from an optimised pulse is moderate in the case of
white amplitude noise considered here, reducing the error in the transfer by a factor
of a little over two compared to the original pulse shape presented by Cirac \etal.\
\cite{cir97}.  Though this is not a very large improvement, regardless
of the shape chosen, pulse shaping will be needed to implement \qst, and so it is prudent to
select the pulse with lowest noise sensitivity.

The order of magnitude of transfer errors produced by the pulse are set by the 
magnitude of $\epsilon_{j}$, which is essentially the relative variance of the 
pulse fluctuations, integrated over time.  As mentioned earlier, for white 
amplitude noise, one candidate that could produce this is intrinsic amplitude 
fluctuations in the laser.  For a cw laser, this can almost immediately be ruled out 
as a legitimate error source, since the relative RMS amplitude lasers in e.g. a diode laser 
are of order $10^{-6}$ \cite{ohtsu}, so the variance is $\epsilon_{j}\sim 10^{-12}$, which is 
completely negligible.

A more likely source of error may be due to either the pulse generator or the pulse
shaper.  Whilst the original proposal for \qst\ was based on atomic states, we have
had in mind the application of this scheme to electron spin states confined to
quantum dots, with the dots contained within a linear microcavity, in a similar
scheme to that illustrated by Benson \etal.\ \cite{ben00} and suggested by Imamo\=glu
\etal.\ \cite{ima99, ima00}.  For a cavity of length 1 \microns, the free spectral
range is around $10^{14}$ Hz, and if we imagine one of the Bragg mirrors has
transmissivity of $10^{-3}$, then the cavity loss rate is around $\kappa\approx
10^{11} \mathrm{sec}^{-1}$.  This is an optimistic but not unreasonable estimate for
the transmissivity, e.g. Lidzey \etal.\ \cite{lid98} report a microcavity $Q$-factor of 125 (based
on a cavity linewidth of $\sim20$ meV and a mode frequency of 2.88 eV) for a
$\lambda/2$ cavity, which corresponds to a mirror transmissivity of
$T_{\mathrm{mir}}=0.025$ \cite{mandelandwolf}, and good quality cavity linewidths may be as low as
0.5 meV \cite{yamamoto}, corresponding to $T_{\mathrm{mir}}=5\times10^{-4}$.

Since $1/\kappa$ sets the time scale of the control pulse (see e.g.
\fig{fig:optimum}), the typical width of the pulse is $\sim 10$ ps.  Weiner
\cite{wei00} gives a relation for the shortest temporal feature that one may generate
as $\delta t=T/\eta$ where $T$ is the largest temporal window, and $\eta$ is a
measure of the potential complexity of the pulse.  Typical values are quoted as
{$T=26.4$ ps}, which is the same order of magnitude as $1/\kappa$, and $\eta=264$,
so {$\delta t\approx10^{-13}$ s}.  Thus, if we assert that fluctuations at time
scales less than $\delta t$ are uncontrolled, then an upper estimate on relative amplitude
fluctuations might be $\delta g\approx g'(t) \delta t$, which means that
uncontrolled relative amplitude fluctuations from the pulse shaper could be as large
as $1/\eta\sim 0.01$, which is certainly not negligible.

We note in passing that the time scale of 10 ps is roughly coincident with the
characteristic time required for an electron in a travelling quantum dot generated by
a surface acoustic wave (SAW) (velocity $\approx 3000$ m s${}^{-1}$), to traverese a
distance of 1 \microns, which is the typical dimension of a SAW wavelength
\cite{bar00}, as well as a typical minimum spot size for an optical CW laser at its focal
point.  Therefore, if a microcavity were constructed at the end of the SAW based quantum
computer proposed in \cite{bar00}, then a CW laser focused at the center of the
microcavity with a cross-sectional intensity profile shaped such that an electron
passing through the microcavity experiences a time varying electric field which
implements \qst\ (e.g \eqn{eq:g1sol}), then the \qst\ scheme discussed in this paper
could be used for both an interconnect between
several SAW quantum processors and for electron-spin measurement (detection of a photon out of the cavity corresponds to a definite spin state of the electron).

This paper has been fundamentally concerned with time dependent Hamiltonians, which
are, conceptually, generally applicable to schemes for implementation of quantum
computation.  We therefore suppose that in many quantum gates implemented with time
dependent control pulses, there could be some advantage in shaping the control pulses to be
minimally sensitive to errors, so the ideas discussed in this paper may well be
generalisable to other quantum information applications.

\section{Summary}

We have analyzed the effect of stochastic noise on a control pulse that implements
\qst\ between two atoms or quantum dots.  We have found an alternate analytic solution for
laser pulses that implement \qst, and using the two analytic solutions we have
available, we found that although different control pulses are capable of
implementing \qst, they may have different sensitivities to noise, depending both on
their shape and the source of the noise.  We then provided a method for calculating
pulse shapes with lower noise sensitivities thereby optimising the pulse shape.  This
method was demonstrated to produce more optimal control pulses for the case of noise
that is proportional to the control pulse amplitude with a white spectrum.  A number
of successively better pulse shapes were presented, and extrapolating the improvement
in the noise sensitivity to its limit (i.e. for infinitely wide pulses) strongly
suggests that the influence of this particular form of noise cannot be cancelled completely.

\begin{acknowledgments}
We would like to thank G.\ Milburn for a very helpful discussion, S.\ Barrett, A.\
Moroz, A.\ Shields, R.\ Phillips, P.\ Littlewood and P.\ Haynes for fruitful conversations regarding implementation of the ideas presented.  TMS acknowledges support from the Hackett Studentship and UK ORS Awards Scheme.  CHWB thanks the EPSRC for financial support.
\end{acknowledgments}

\end{document}